\documentclass[%
 pre, onecolumn,
superscriptaddress,
 amsmath,amssymb,
 aps,
nofootinbib
]{revtex4-2}

\usepackage{graphicx}
\usepackage{dcolumn}

\usepackage{bm}

\usepackage{xcolor,soul}
\usepackage{xr}
\usepackage{comment}
\usepackage{hyperref}
\usepackage{cleveref}
\usepackage{enumitem}
\usepackage{lineno}
\usepackage{accents}
\hypersetup{colorlinks,linkcolor={blue!90!black},citecolor={blue!90!black},urlcolor={blue!90!black}}

\DeclareMathAlphabet\mathbfcal{OMS}{cmsy}{b}{n}

\usepackage{amsmath}
\usepackage{mathtools}

\begin{document}
\title{Feeders and Expellers, Two Types of Animalcules With Outboard Cilia,\\
Have Distinct Surface Interactions}
\author{Praneet Prakash }
\email[]{pp467@cam.ac.uk}
\affiliation{Department of Applied Mathematics and Theoretical 
Physics, Centre for Mathematical Sciences,\\ University of Cambridge, Wilberforce Road, Cambridge CB3 0WA, 
United Kingdom}
\author{Marco Vona}
\email[]{mfv25@cam.ac.uk}
\affiliation{Department of Applied Mathematics and Theoretical 
Physics, Centre for Mathematical Sciences,\\ University of Cambridge, Wilberforce Road, Cambridge CB3 0WA, 
United Kingdom}%
\author{Raymond E. Goldstein}
\email[]{R.E.Goldstein@damtp.cam.ac.uk}
\affiliation{Department of Applied Mathematics and Theoretical 
Physics, Centre for Mathematical Sciences,\\ University of Cambridge, Wilberforce Road, Cambridge CB3 0WA, 
United Kingdom}%

\date{\today}

\begin{abstract}
 Within biological fluid dynamics, it is conventional to distinguish between 
``puller" and ``pusher" microswimmers on the basis of the forward or aft 
location of the flagella relative to the cell body: typically, bacteria are
pushers and algae are pullers.  Here we note that since many pullers have 
``outboard" cilia or flagella displaced laterally from the cell centerline on both
sides of the organism, there are two important subclasses whose
far-field is that of a stresslet, but whose near field is qualitatively more complex.  
The ciliary beat creates not only a propulsive force
but also swirling flows that can be represented by paired rotlets with two possible
senses of rotation, either ``feeders" that sweep fluid toward the cell apex, or 
``expellers" that push fluid away.  Experimental studies of the 
rotifer \textit{Brachionus plicatilis} in combination with earlier work on the green algae 
\textit{Chlamydomonas reinhardtii} show that the two classes have markedly
different interactions with surfaces.  When swimming near a surface, expellers such as 
\textit{C. reinhardtii} scatter from the wall, 
whereas a feeder like
{\it B. plicatilis} stably attaches.  This results in a stochastic ``run-and-stick" locomotion, with periods of 
ballistic motion parallel to the surface interrupted by trapping at the surface.
\end{abstract}
\maketitle

\section{Introduction}
\label{intro}

In the description of both individual and collective dynamics
of motile microorganisms a considerable simplification can
often be achieved by partitioning their effect on the
surrounding fluid into separate contributions from the
organism body and the appendages \textemdash cilia or flagella\textemdash that confer locomotion.
These contributions can further be simplified 
into those of equal and opposite point forces acting on the 
fluid, as required by the force-free condition on a free swimmer.  Thus, peritrichously 
flagellated bacteria, with a bundle of rotating flagella aft of the body, are termed 
``pushers", while the 
breast-stroke beating of paired algal flagella forward of the body defines a ``puller" 
\cite{SaintillanShelley,Laugabook}.  Direct measurements of the flow fields around freely-swimming 
algae \cite{Direct} and bacteria \cite{Noise} have confirmed that the far-field flows
are consistent with the singularity representation of swimmers.

The force dipole picture gives considerable insight into
many features of swimming.  Viewing 
microswimmer suspensions as a collection of 
interacting stresslets leads to an understanding \cite{Ramaswamy,SaintillanShelley} 
of why a bacterial suspension exhibits ``bacterial turbulence" 
\cite{Dombrowski,Mesoscale,BacTurb} while a suspension of algae does 
not.  At the individual level, the singularity
picture explains accumulation of sperm 
cells at no-slip surfaces due to the reorientation of pushers to become parallel to such 
boundaries \cite{Berke}. 
Attractive interactions between Stokeslets near a no-slip surface \cite{Blake,Squires} underlie
the formation of ``hydrodynamic bound states" of \textit{Volvox} colonies \cite{Dancing}, 
in which negatively buoyant chiral microswimmers are attracted together in the plane of the
surface and orbit each other, a phenomenon later seen in several other systems \cite{Petroff,Pierce,Tan,Modin}.

\begin{figure*}[t]
\centering
\includegraphics[width=1.0\columnwidth]{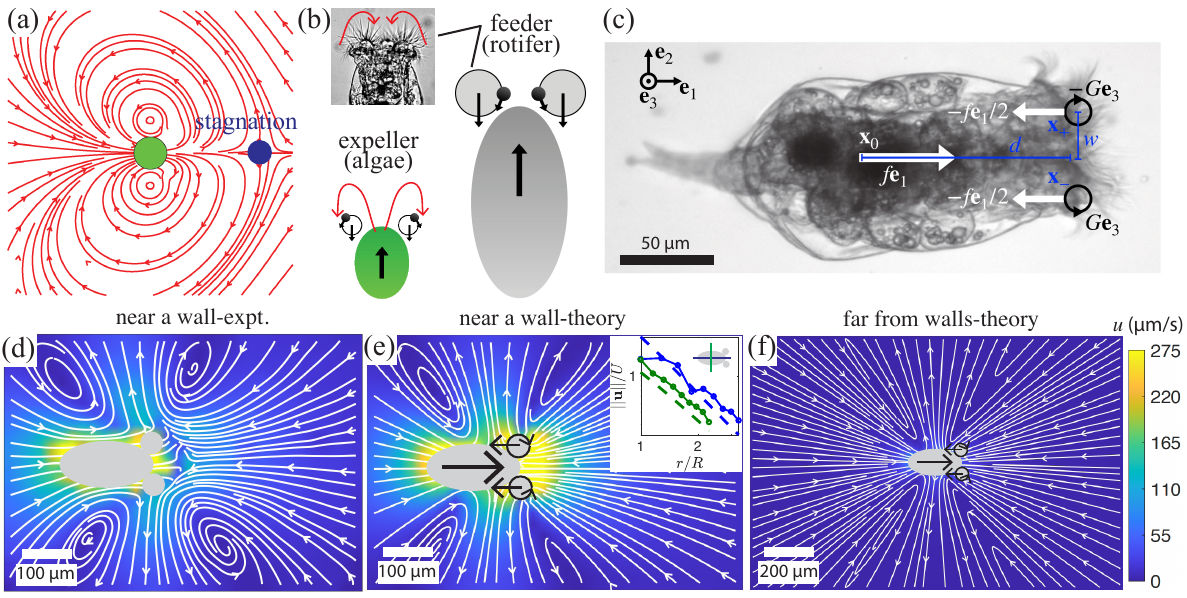}
\caption{Flow fields of expellers and feeders. 
(a) Illustrative theoretical flow field for \textit{Chlamydomonas 
reinhardtii} using a superposition of three Stokeslets and 
two rotlets (not fitted). (b) Schematics showing outward ciliary movement of the 
expeller and inward ciliary movement of the feeder. (c) Schematic representation of the body material frame and singularities employed in 
\eqref{FittingAnsatz}. The body frame consists of $\mathbf e_1$ (aligned with the body axis), 
$\mathbf e_2$ (in the plane of the coronae, pointing in either direction) and 
$\mathbf e_3=\mathbf e_1\times\mathbf e_2$ (pointing out of the page). The effect of the 
body on the fluid is represented by means of a system of Stokeslets and rotlets. The body 
Stokeslet of strength $f\mathbf e_1$, representing the drag on the fluid, is balanced by 
two $-f\mathbf e_1/2$ Stokeslets at $\mathbf x_{\pm}=\mathbf x_0+d\mathbf e_1\pm w\mathbf e_2$. 
The circulating flow induced by the cilia is modelled as two rotlets of strength $\mp G\mathbf e_3$ 
placed at $\mathbf x_{\pm}$.
(d) Magnitude and streamlines of the rotifer
flow field. (e) Fitted 
approximation to (d), along with $\log$-$\log$ plot of the velocity decay along (blue) and 
perpendicular to (green) the body axis; arrows schematically denote the orientation of 
the Stokeslets and rotlets. (f) Magnitude and streamlines of theoretical flow field for 
a rotifer a distance from the wall $10^4$ times larger than its fitted value in (e); the 
topology of the flow changes with only the trailing stagnation points surviving
\cite{ConfinedStokeslet}.}
\label{fig1}
\end{figure*}

Not surprisingly, the force dipole representation alone may fail to capture near-field effects, 
which may require higher-order singularities or the invocation of lubrication forces
which become important in the near-field \cite{Spagnolie}, an effect documented in swimming 
spermatozoa \cite{QuadrupoleInteractions} and {\it E.~coli} \cite{EColiQuadrupole}.
A clear breakdown of the stresslet picture is provided by 
interactions of the unicellular green alga \textit{Chlamydomonas reinhardtii} with 
surfaces \cite{Kantsler}.
Whereas the stresslet approximation predicts that pullers nosedive into no-slip 
surfaces, experiments show instead an 
``inelastic scattering" phenomenon, where almost all incoming angles of swimming trajectories 
lead to approximately zero outgoing angle, corresponding to swimming parallel to the surface.  
In these experiments, the reorientation at the surface
was shown to arise from direct ciliary contact interactions.  Important later work
\cite{PolinScattering} on scattering of {\it Chlamydomonas} by curved no-slip surfaces showed
that similar geometry of reorientation can arise from hydrodynamic 
interactions without the need for direct contact with a wall.

The hydrodynamic interactions responsible for the rotation of algae away from 
the perpendicular orientation favored by the puller stresslet arise from the 
the undulatory beating of the two algal cilia.
We term these ``outboard" cilia, as they are displaced laterally on either side of the
cell centerline.  The time-averaged flow field around {\it Chlamydomonas} \cite{Direct},
shown in Fig. \ref{fig1}(a), as well as the time-resolved flow field \cite{Guasto}, 
illustrates the swirling action of the flagellar beat, with
an extended power stroke driving flows backward, and a contracted recovery stroke 
nearer the cell body, driving weaker flows forward.  These complex 
time-averaged fields
can be  
represented by a superposition of three Stokeslets: one for the cell body
pointing forward, and one in the middle of each flagellum, pointing rearward, as in Fig. \ref{fig1}(b).
The fully time-dependent problem can be modelled by time-varying combinations 
of singularities \cite{Quadroar,Klindt,ChlamyScattering}.

The notion that the flow field arising from the beating strokes of 
eukaryotic cilia can be represented by a Stokeslet appears as well in studies of 
so-called ``mosaic" ciliated tissues \cite{Mosaic}, such as the epidermis of developing amphibians,
in which a sparse population of multiciliated cells exists in a background of nonciliated cells.
In the simplest picture, the action of a large number of cilia, with no phase synchrony, in 
driving flow along the tissue surface can be quantitatively captured by a single
point force parallel to the surface.    The representation of the flow due to a flagellum as
that of a moving Stokeslet also forms the basis for a very large
amount of work on the synchronization of cilia \cite{NEL}. 
Yet, it is also intuitively reasonable that the 
orbits of the flagella, with the extended power stroke and contracted recovery stroke, could also
be modelled as a point torque on the fluid, or a rotlet.
Importantly, the flow fields
for parallel orientation of both Stokeslets and rotlets near a no-slip surface both decay
as $1/r^2$ \cite{Blake,BlakeChwang}, so there is no way from the far-field decay alone to prefer 
one representation over another.  Recent work has, however, shown that 
the time-averaged flow generated by a beating cilium near a wall is better 
represented as a rotlet, especially for near-field transport \cite{RotletCilia}.

Here we reconsider the problem of microswimmers with outboard cilia in light of the background
summarized above.  Our primary observation is that Nature presents us with two broad classes of
such organisms, distinguished by the direction of swirling flows created by the cilia.  As 
in Fig. \ref{fig1}(a), the breaststroke beating of biflagellates such as 
{\it Chlamydomonas} sweeps fluid {\it away} from the cell apex; we name these ``expellers".
By contrast, the corona of cilia in more complex multicellular organisms such as the rotifer 
\textit{Brachionus plicatilis}
shown in Figs. \ref{fig1}(b,c) directs flow {\it toward} the mouth, and are naturally termed 
as ``feeders".
Rotifers are complex ``animalcules" with 
internal organs and a nervous system, and serve as model organisms for a wide 
range of biological processes, from evolution \cite{rotifers_evolution} 
to aging \cite{rotifers_aging}.  
The first observation of rotifers is variously attributed \cite{Ford} to Antony van Leeuwenhoek 
\cite{vL1674} and John Harris \cite{Harris1696}, with decisive descriptions due to
the former in a series of papers in the early 
years of the 18$^{\rm th}$ century \cite{vL1703,vL1704,vL1705}. Even in these very early works
there is reference made to the tendency of rotifers to attach strongly to surfaces, which 
in light of the observations of surface scattering of \textit{Chlamydomonas} serves
to illustrate the fundamental distinction in surface interactions between expellers and feeders.

After outlining in Sec. \ref{methods} the experimental methods used here to study the 
swimming and surface interactions of rotifers, we present in Sec. \ref{flowfields}
a quantitative analysis of the flow field around a freely-swimming rotifer and its representation
in terms of a superposition of Stokeslets and rotlets.  This leads naturally to consideration of
the interactions of rotifers with no-slip surfaces, considered in Sec. \ref{scattering}, where
we show that they exhibit a rapid transition from freely swimming to surface attachment.
This can be understood quantitatively through a model akin to the stresslet one in which the additional contribution from
the outboard cilia is represented by a rotlet doublet.
The linear stability problem of such ``composite" swimmers near a surface is studied in Sec. \ref{linstab}.
Finally, Sec. \ref{motility} examines trajectories on larger spatial and temporal scales.  We show experimentally that rotifers swimming
near a surface exhibit the phenomenon of ``run-and-stick", in which roughly straight line swimming
is interrupted stochastically by trapping at the surface through the mechanism discussed in Sec. \ref{linstab}.
A simple model of stochastic switching between bound and free states, similar in spirit to one
used to study analogous transitions in \textit{E. coli} \cite{Peruani}, is shown to capture the
essential features of the observations.   The concluding Sec. \ref{discussion} highlights possible
future directions of this research.  Various details of calculations and data analysis are
collected in Appendices \ref{app_a}-\ref{app_d}

\section{Experimental Methods}
\label{methods}

We use the rotifer \textit{Brachionus plicatilis} (strain 5010/4) as a model feeder 
organism \cite{Fussman2000}. It is approximately $210\,\mu$m in length and $90\,\mu$m in width, 
with individual cilia of
length $\sim 50\,\mu$m and beat frequency $\sim20-30$\,Hz. The cells can attain swimming speeds 
of $200-400\,\mu$m/s. 
They were grown in a coculture with the alga
\textit{Dunaliella tertiolecta} (strain 19/7c) as a food source in marine f/2 medium at 
$20\,^{\circ}$C, under a diurnal cycle of 12 h cool white light ($\sim 15 \, \mu$mol photons/m$^2$s 
PAR) and $12\,$h in the dark. All strains and media concentrates were 
sourced from the Culture Collection of 
Algae and Protozoa (CCAP) \cite{CCAP}. To isolate \textit{B. plicatilis}, the coculture 
was passed through a $70\,\mu$m diameter membrane filter (pluriStrainer) to remove the algae. 

\textit{B. plicatilis} is an invertebrate, with a dense ciliary array at the cell apex and a 
tail at the posterior end as shown in Fig. \ref{fig1}(c). While swimming, it retracts 
its tail, appearing like a prolate ellipsoid, and the cilia in front organize into two clusters 
of biaxial symmetric metachronal bands on either side of the cell axis, sweeping fluid 
towards the centrally located mouth. To quantify flow fields, we acquired 
brightfield images of \textit{B. plicatilis} in f/2 medium infused with polystyrene tracer 
particles (mass fraction {$0.01$\,\%}). Images were captured at 500 fps using a $\times 10$ 
objective with a high-speed camera (Phantom V311) mounted on a Nikon TE-2000U inverted 
microscope. A dilute suspension of rotifers ($100$ cells cm$^{-3}$) mixed with tracer particles was 
transferred into a chamber formed by two coverslips separated by a double-sided tape of 
thickness $4\,$mm. Fluid velocimetry using PIVlab was performed to analyze the flows 
\cite{PIVLab}. The rigid boundary helps the rotifer swim parallel near to the surface long enough to 
capture high-resolution images of its flow-field. When far from the surface, image capture is 
difficult because the swimmers frequently move out of the focal plane, causing their bodies to 
appear defocused. 

\section{Flow fields}
\label{flowfields}

Figure \ref{fig1}(d) shows time-averaged experimental flow-field 
of \textit{B. plicatilis}  swimming at a speed of 
$380\ \mu\text{m}\ \text{s}^{-1}$ nearly 
$350\,\mu$m away from the upper surface of the chamber (See also SM Video 1 \cite{SM}).
Assuming a typical rotifer size of $L=210\ \mu$m, 
a maximum swimming speed of $U=400\,\mu$m/s, and a 
kinematic viscosity of $\nu=10^6\ \mu$m$^2$/s, 
the Reynolds number does not exceed $\text{Re}=UL/\nu\sim 0.08$. 
We therefore work in the inertia-free limit $\text{Re}=0$. We endow the
swimmer centre of mass $\mathbf x_0$ with an orthonormal body-fixed frame 
$\{\mathbf e_1,\mathbf e_2,\mathbf e_3\}$, where 
$\mathbf e_1$ is the swimming direction and $\mathbf e_2$ is 
the normalised displacement between the cilia bundles located at 
\begin{equation}
\mathbf x_{\pm}=\mathbf x_0+d\mathbf e_1\pm w\mathbf e_2,
\end{equation}
as in Fig.~\ref{fig1}(c). 
The experimental flow in Fig.~\ref{fig1}(d) 
was fit to a superposition of three Stokeslets of strengths $f\mathbf e_1$ 
(body drag), $-f\mathbf e_1/2$, $-f\mathbf e_1/2$ (thrust) located at $\mathbf x_0$,
$\mathbf x_{\pm}$, as well as two rotlets with strengths 
$\mp G\mathbf e_3$ at $\mathbf x_{\pm}$, representing the 
sweeping ciliary flow towards the mouth \cite{Spagnolie, ChlamyScattering} (Fig.~\ref{fig1}(c) and Appendix
\ref{app_a}),
\begin{equation}
\mathbf u_{\text{fit}}(\mathbf x)=\frac{f}{8\pi\mu}\left[\mathbf B(\mathbf x;\mathbf x_0)
-\frac{1}{2}\mathbf B(\mathbf x;\mathbf x_+)-\frac{1}{2}\mathbf B(\mathbf x;\mathbf x_-)\right]\cdot\mathbf{e}_{1}
+\frac{G}{8\pi\mu}\left[\mathbf R(\mathbf x;\mathbf x_-)-\mathbf R(\mathbf x;\mathbf x_+)\right]\cdot\mathbf e_3.
\label{FittingAnsatz}  
\end{equation}
Here $\mathbf B$ and $\mathbf R$ are respectively the Blake tensors for a point force and torque near a 
no-slip wall \cite{Blake}.
The fit yields $f\sim 400\ \text{pN}$, $G\sim 4\ \text{pN mm}$, $U\sim 400\ \mu\text{m}\ \text{s}^{-1}$, 
$d\sim 130\ \mu\text{m}$, $w\sim 40\ \mu\text{m}$. The rotlets displacements $d$ and $w$ are 
comparable to the semi-axes, reflecting the bundles' locations. Furthermore, for a bundle size of 
$25-50$ cilia, the force per cilium is on the order of $4-10$ pN, consistently with the literature 
\cite{KyriacosPhototaxis}. A positive value of $G$ makes the organism a ``feeder'', as in 
Fig.~\ref{fig1}(b). The torque exerted by a single bundle should be on the order of 
$(\text{thrust per cilium})\times (\text{bundle circumference})$. Assuming a thrust of $10$ pN per cilium 
and a bundle radius of $25\ \mu$m, the estimate $G\sim 2$ pN mm resembles the fitted value. 
Finally, neglecting the no-slip wall and treating the body as a prolate ellipsoid, the swimming speed 
in an unbounded fluid is related to the thrust by $f=6\pi\mu b\zeta U$, where \cite{HappelBrenner}
\begin{equation}
\zeta=\left\{\frac{3}{4}(\xi^2-1)^{1/2}[(\xi^2+1)\text{arcoth}(\xi)-\xi]\right\}^{-1}, 
\ \ \ \ {\rm and} \ \ \ \ \xi=\frac{a}{(a^2-b^2)^{1/2}}.  
\end{equation}
Taking the semi-axes to be $a\sim 100\ \mu$m, $b\sim 50\ \mu$m and assuming $\mu=10^{-3}$ Pa s, a thrust of 
$400$ pN should result in a swimming speed $U\sim 350\ \mu\text{m}\ \text{s}^{-1}$, 
similar to the experimental value. The fitted flow in Fig.~\ref{fig1}(e) displays good quantitative 
agreement with the experimental data in Fig.~\ref{fig1}(d) up to $1.5$ body lengths ahead of the swimmer. 
The flow topology of rotifers changes far from the chamber wall, where the two elliptic stagnation points 
forward of the bundles Fig.~\ref{fig1}(d,e) disappear leaving only the trailing stagnation points 
\cite{ConfinedStokeslet,Noise,Lobes}. Notably, unlike for \textit{C. reinhardtii} in Fig.~\ref{fig1}(a) 
\cite{Direct}, 
the rotifer flow field in the absence of a boundary (Fig.~\ref{fig1}(f)) lacks a stagnation point ahead 
of the body, owing to the smaller aspect ratio and the opposite circulation near the cilia bundles.
The inset in Fig.~\ref{fig1}(e) suggests a decay rate $r^{-1.43}$ for the experimental flow perpendicular 
to the body axis and a $r^{-1.78}$ decay along it. These are consistent with $r^{-2}$ stresslet flow, 
which is unaffected by the presence of the wall within a distance $r\ll 2h\sim 720\ \mu$m of the 
singularities, with $h$ being the fitted height above the chamber wall. 
For $r\gg 2h$, the presence of the wall is felt and the flow decays like $r^{-3}$, rather than $r^{-2}$.  Interestingly, very strong confinement by two no-slip surfaces has been observed to reverse the sense of circulation of the vortices straddling the body of \textit{Chlamydomonas}, turning an expeller flow into one that resembles a feeder \cite{Mondal}.

\section{Scattering From a No-slip Surface}
\label{scattering}

While the interactions between pusher swimmers and boundaries have been shown to be chiefly 
hydrodynamic \cite{Berke}, as discussed in the introduction, 
there is evidence that pullers can interact with walls both 
through direct wall contact \cite{Kantsler,PolinScattering} and through
the fluid \cite{ChlamyScattering, PolinScattering}. Moreover, while the leading-order stresslet 
flow suggests that pullers should nosedive into the wall, both scattering 
\cite{PolinScattering,Kantsler} and trapping are observed experimentally.
Here we study the hydrodynamic picture of the different scattering properties 
of feeders and expellers, exemplified by \textit{B. plicatilis} and \textit{C. reinhardtii}. 
We employ a far-field expansion that 
includes both the force dipole term \cite{PedleyKessler1992}, and the rotlet dipole 
that arises from contracting the two opposite rotlets into a single singularity.
This approach offers a simple framework that can be incorporated into simulations of many-swimmer
systems and coarse-grained continuum theories \cite{SaintillanShelley}. 

We quantify the strength of the rotlets via the signed ``feeder number'' 
\begin{equation}
    \text{Fe}=\frac{Gw}{fd L},
\end{equation}
where $L$ is a again the typical swimmer size.  We shall refer to swimmers with $|\text{Fe}|\gg (\ll) 1$ as 
``strong (weak) feeders/expellers''. \textit{B. plicatilis} is a weak 
feeder ($\text{Fe}\sim 0.06$) while \textit{C. reinhardtii} is a strong expellers with $\text{Fe}\sim -5$ 
(using $L\sim 5\ \mu$m, $d\sim 5\ \mu$m, $w\sim 10\ \mu$m, $f\sim\ 7.2$ pN, $G\sim -87\ \text{pN}\ \mu$m, 
$U\sim 100\ \mu \text{m}\ \text{s}^{-1}$, and $e\sim 0.75$ \cite{Synchronisation, Direct, KyriacosPhototaxis}). 
The rotlet flow becomes comparable to that of the force dipole at the ``feeder length'' 
\begin{equation}
    \ell=\text{Fe}L.
\end{equation} Indeed, 
$|\ell| \sim 5L$ matches the \textit{C.~reinhardtii} stagnation point \cite{Direct}, located $\sim 6$ radii 
ahead of the cell body.

In the far field $r\coloneqq \lVert\mathbf x\rVert\gg L$, a swimmer may be described as a sum of flow 
singularities, specifically a force dipole of magnitude ${\mathcal O}(fd/\mu r^2)$, 
and higher-order gradients of Stokeslets of 
strength ${\mathcal O}(fd^2/\mu r^3)$ or ${\mathcal O}(fw^2/\mu r^3)$ \cite{Direct,Spagnolie}, 
a rotlet dipole of strength ${\mathcal O(}Gw/\mu r^3)$ and asymptotically smaller terms that we neglect.
Since $d,w\sim L$, the force dipole dominates the quadrupole for $r\gg L$. We note here that if $G\sim fL$ 
as we may expect, then the rotlet dipole ${\mathcal O}(Gw/\mu r^3)\sim {\mathcal O}(f Lw/\mu r^3)$ 
is dominated by the force dipole for $r\gtrsim L$. In practice, however, the relation $G\sim fL$ only applies to 
the peak force and torque $f^*$, $G^*$. Because only a fraction of the peak force is converted into 
thrust ($0.25$ for \textit{C.~reinhardtii}), the average force $f$ which sets the swimming speed 
\cite{KyriacosPhototaxis} is typically much lower ($f\sim 0.25 f^*$ for \textit{C.~reinhardtii}). 
Therefore, while $G^*w/f^*dL\sim -2$, the ``effective'' feeder number is $\text{Fe}\sim -5$ for 
\textit{C.~reinhardtii}. For strong feeders and expellers the dipole is therefore comparable 
to the rotlet dipoles 
for $1\ll r/L\ll |\text{Fe}|$. This validates representing the swimmer as a force dipole and torque dipole term 
of respective strengths $fd$ and $Gw$ centred at the same point $\mathbf x_0$. Despite the 
singularity approximation breaking down near a wall, it has been shown to qualitatively reproduce 
near-field dynamics \cite{Spagnolie,Berke,Noise,Colloid Trapping}. 

The wall modifies the trajectory of the swimmer through an image flow $\mathbf u^*$ ensuring no-slip. 
The translation of a prolate ellipsoid is governed by the Fax{\'e}n equation,
\begin{equation}\dot{\mathbf x}_0=\mathbf 
u^*(\mathbf x_0)+U\mathbf e_1+\mathcal O(L^2\nabla^2\mathbf u^*),
\label{Faxen}
\end{equation}
while the orientational dynamics is expressed by Jeffery's equation \cite{Spagnolie},
\begin{equation}
\dot{\mathbf e}_1=\mathbf{\Omega}\times\mathbf e_1, \ \ \ \\ \mathbf\Omega=
\frac{1}{2}\nabla\times \mathbf u^*(\mathbf x_0)+\Gamma\mathbf e_1\times\mathbf E^*(\mathbf x_0)
\cdot\mathbf e_1+\mathcal O(L^2\nabla^2\nabla\times\mathbf u^*).
\label{Jeffery}
\end{equation}  
Here, 
$\mathbf E^*=(\nabla\mathbf u^*+\nabla^{\text T}\mathbf u^*)/2$ is the image rate-of-strain tensor, 
and $\Gamma=(1-e^2)/(1+e^2)$ is the shear-alignment parameter with $e=b/a$ the aspect ratio. 
We henceforth neglect the asymptotically sub-leading terms in \eqref{Faxen} and \eqref{Jeffery}. 

\begin{figure*}[t]
\centering
\includegraphics[width=0.94\columnwidth]{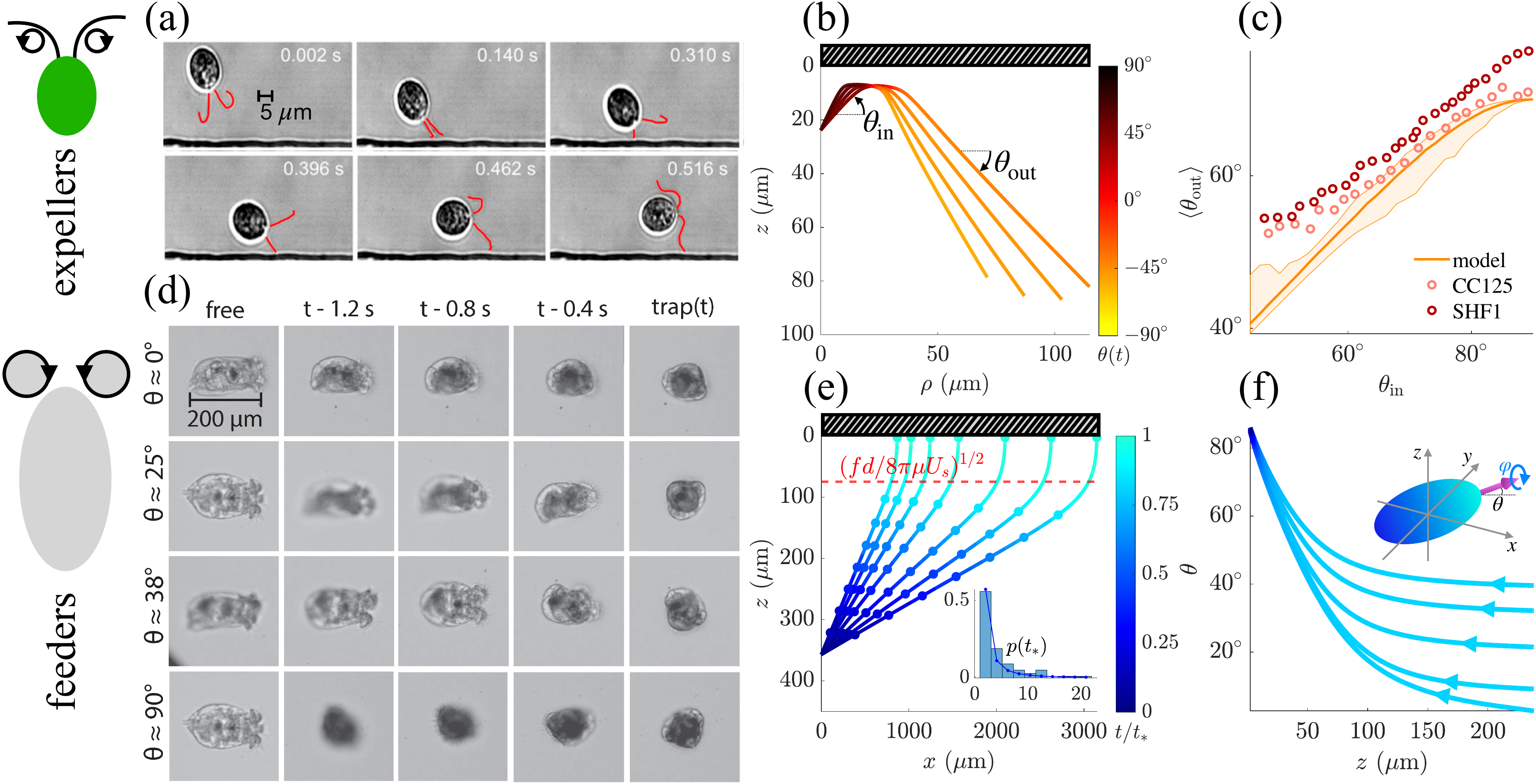}
\caption{Scattering of expellers and feeders from surfaces. (a) Scattering of \textit{C. reinhardtii} off 
a rigid boundary via steric interactions, from \cite{Kantsler} with permission. (b) 
Computed trajectories appropriate to  
\textit{C. reinhardtii}.  
(c) Average scattering angle $\langle\theta_{\text{out}}\rangle$ as a function of 
incident angle $\theta_{\text{in}}$ for \textit{C. reinhardtii}. Orange 
curve shows the predictions of singularity model, with shaded region indicating
variations associated with roll angle.
Circles are the scattering data for the mutants CC125 and SHF1 from \cite{PolinScattering}. 
(d) Trapping of four rotifers with different initial pitches (see also SM Video 2 \cite{SM}). 
(e) Theoretical trajectories of rotifers show ``snapping" to the boundary for a range of initial 
pitches and zero initial roll. The colorbar represents the proportion of the total time elapsed, 
while circles mark the position at regular time intervals, showing a speed increase near the boundary. 
The red line marks the distance $z^*$ in \eqref{zstar} at which the dipole flow becomes 
comparable to the swimming speed. The inset shows the distribution of the impact time $t_*$ and a fitted `ballistic' p.d.f. $p_b(t)=2t_0/\pi t(t^2-t_0^2)^{1/2}$. (f) Evolution of the 
pitch and roll $\theta$ and $\phi$ for five trajectories with different $\theta_0$ and $\phi_0$. 
Variations in the roll are correlated with the color shade. As the distance to the wall decreases,
the pitch increases to $90^{\circ}$ and the roll changes abruptly near impact.}
\label{fig2}
\end{figure*}

We now apply the above to the scattering of microswimmers from a no-slip surface, focusing first
on expeller dynamics exemplified by \textit{Chlamydomonas}.  For context, 
Fig. \ref{fig2}(a) shows a typical scattering event of \textit{C. reinhardtii} in which
one cilium makes brief contact with the surface.  Figure \ref{fig2}(b) examines the case in which 
purely hydrodynamic interactions govern scattering, obtained by numerical integration of the
dynamics \eqref{Faxen} and \eqref{Jeffery}, 
using typical values $d\sim 5\ \mu$m, $w\sim 10\ \mu$m, $f\sim\ 7.2$ pN, $G\sim -87\ 
\text{pN}\ \mu$m, $U\sim 100\ \mu \text{m}\ \text{s}^{-1}$, $e\sim 0.75$ 
\cite{Synchronisation,Direct,KyriacosPhototaxis}.
For a range of incoming angles 
$\theta_{\text{in}}$ we observe gliding along the wall for a short distance followed by
turning away from the wall with a scattering 
angle $\theta_{\text{out}}$ that is monotonically increasing with the incident angle. 
Figure \ref{fig2}(c) plots the scattering data of Contino, et al. \cite{PolinScattering}
for the wild type strain CC125 and the short flagella mutant SHF1 of \textit{C. reinhardtii} 
in the range $\theta_{\text{in}}\gtrsim 44^{\circ}$ where boundary interactions are 
mostly hydrodynamic.  In order to compare the results of the singularity model with these data
we computed the  
$\langle \theta_{\text{out}}\rangle$ for $10^3$ random values of the roll angle.   The 
average value and spread associated with the roll angle are shown respectively as a solid line
and shaded region in Fig. \ref{fig2}(c) indicate
that this model
accurately reproduces the approximately linear growth of $\langle\theta_{\text{out}}\rangle$ 
with $\theta_{\text{in}}$. Indeed, we fit $\langle\theta_{\text{out}}\rangle
\sim 0.69 \cdot\theta_{\text{in}}+11.13^{\circ}$, while $\langle\theta_{\text{out}}\rangle\sim 0.59 
\cdot\theta_{\text{in}}+22^{\circ}$ for the CC125 mutant and $\langle\theta_{\text{out}}\rangle\sim 
0.64\cdot\theta_{\text{in}}+21^{\circ}$ for the SHF1 mutant \cite{PolinScattering}. These results
provide a quantitative validation of the model of \textit{Chlamydomonas} as a swimmer governed by
the sum of a puller stresslet and an expeller rotlet doublet.

Turning to rotifers, Fig. \ref{fig2}(d) shows examples of their rapid 
``snapping" to a solid boundary.
In the geometry of an inverted microscope, these are views from below as rotifers that are initially 
oriented with their long axis $\mathbf e_1$ parallel to the upper chamber coverslip turn and rapidly rotate to become 
attached to that 
surface, such that $\mathbf e_1$ points away from the observer.  As remarked earlier, this
phenomenon was known to van Leeuwenhoek, who said in his famous letter to the Royal Society of 
17 October, 1687, \textit{``... These Animals also had a 
second movement; for when they were unable to make any progress by swimming, they attached themselves to
the glass by the organs at the front of the head; and then they drew their body up short..."} 
\cite{vL1687}.  Our studies show that the process of snapping typically 
takes on the order of $1-2\,$s from initiation to vertical alignment.
Once attached, rotifers are observed to spin around $\mathbf e_1$ with a period of $\sim 5\,$s, 
sometimes for many complete rotations before ultimately detaching.  This motion is likely
related to the spinning motion around $\mathbf e_1$ seen during free swimming. 
We discuss the long-time statistics 
of the switching between swimming and sticking in Sec. \ref{motility} below.

Using the fitted values of $fd$ and $Gw$ in the stresslet+rotlet doublet model, Fig. \ref{fig2}(e) 
shows that numerically obtained snapping trajectories recapitulate the behavior observed in 
Fig.~\ref{fig2}(d).  As the wall is approached the flow due to the strong puller stresslet 
increases as $1/z^2$, and thus we can define the lengthscale 
\begin{equation}
z^*=\left(\frac{fd}{8\pi\mu U}\right)^{1/2}
\label{zstar}
\end{equation} 
at which the stresslet flow is comparable to the swimming speed.  From the angular dynamics in
Fig.~\ref{fig2}(f) we deduce that the body turning is driven by the strong puller stresslet.
Defining the yaw $\chi$, pitch $\theta$, and roll $\varphi$ of the body frame by 
$\{\mathbf e_1,\mathbf e_2,\mathbf e_3\}=R_z(\chi)R_y(\theta)R_x(\varphi)$, 
Fig.~\ref{fig2}(f) shows that the pitch rapidly increases to nearly $90^{\circ}$ and impact is 
associated with abrupt twisting motion along the body axis.

\section{Stability Analysis of Trajectories}
\label{linstab}

In this section we show that the presence of ``feeder'' rotlet flow turns a swimmer 
located at a distance $h$ from a wall towards the
no-slip wall when the trajectory is nearly perpendicular or parallel to the wall. Conversely, 
an ``expeller flow'' generally rotates the rotifer away from the wall. For a 
trajectory that is nearly perpendicular to the wall, i.e.~with $\mathbf e_1=-\hat{\mathbf z}
-\varepsilon \mathbf V\times \hat{\mathbf z}$, $\mathbf e_2=\hat{\mathbf y}
+\varepsilon \mathbf V\times\hat{\mathbf y}$, $\mathbf e_3=\hat{\mathbf x}+\varepsilon 
\mathbf V\times\hat{\mathbf x}$, up to $\mathcal O(\varepsilon^2)$, a careful 
analysis given in Appendix \ref{app_c} shows that 
the swimmer moves towards the wall with speed
\begin{align} 
\dot h=-\left[U+\frac{3fd}{32\pi \mu h^2}\left(1+\frac{4\text{Fe}L}{3h}\right)\right]+\mathcal O(\varepsilon)
\end{align}
Therefore, feeders speed up as they approach the wall, while expellers slow down or even hover if $\dot h\sim 0$. 
Rotlets control the snapping if the crossover length $h\sim \text{Fe}L =\ell$ 
is much larger than the body size, i.e.~if the organism is a strong feeder or expeller. 
Moreover, a nosediving puller is instantaneously rotated towards the wall if
\begin{align}
(7\Gamma-1)\frac{\ell}{2h}+\Gamma+1>0, \ \ \ \ \  \frac{5\ell}{2h}+1>0.
\end{align}
For large force dipoles ($|\ell/h|\ll 1$) the swimmer always aligns perpendicular to the 
wall \cite{Spagnolie}, while for strong rotlet dipoles or near collision ($|\ell/h|\gg 1$) the 
swimmer tends to align parallel to the wall when $G<0$ (expeller) and perpendicular to the wall when 
$G>0$ (strong feeder) provided the body is not too spherical. This analysis reveals a fundamental 
difference between rotifers and green algae: whereas the former behave essentially like 
stresslets when colliding with the wall (and are thus stable), expeller algae feel the 
destabilizing effects of the rotlet and are thus unstable [see Fig.~\ref{fig2}(b,c)].

\begin{figure*}[t]
\centering
\includegraphics[width=0.7\columnwidth]{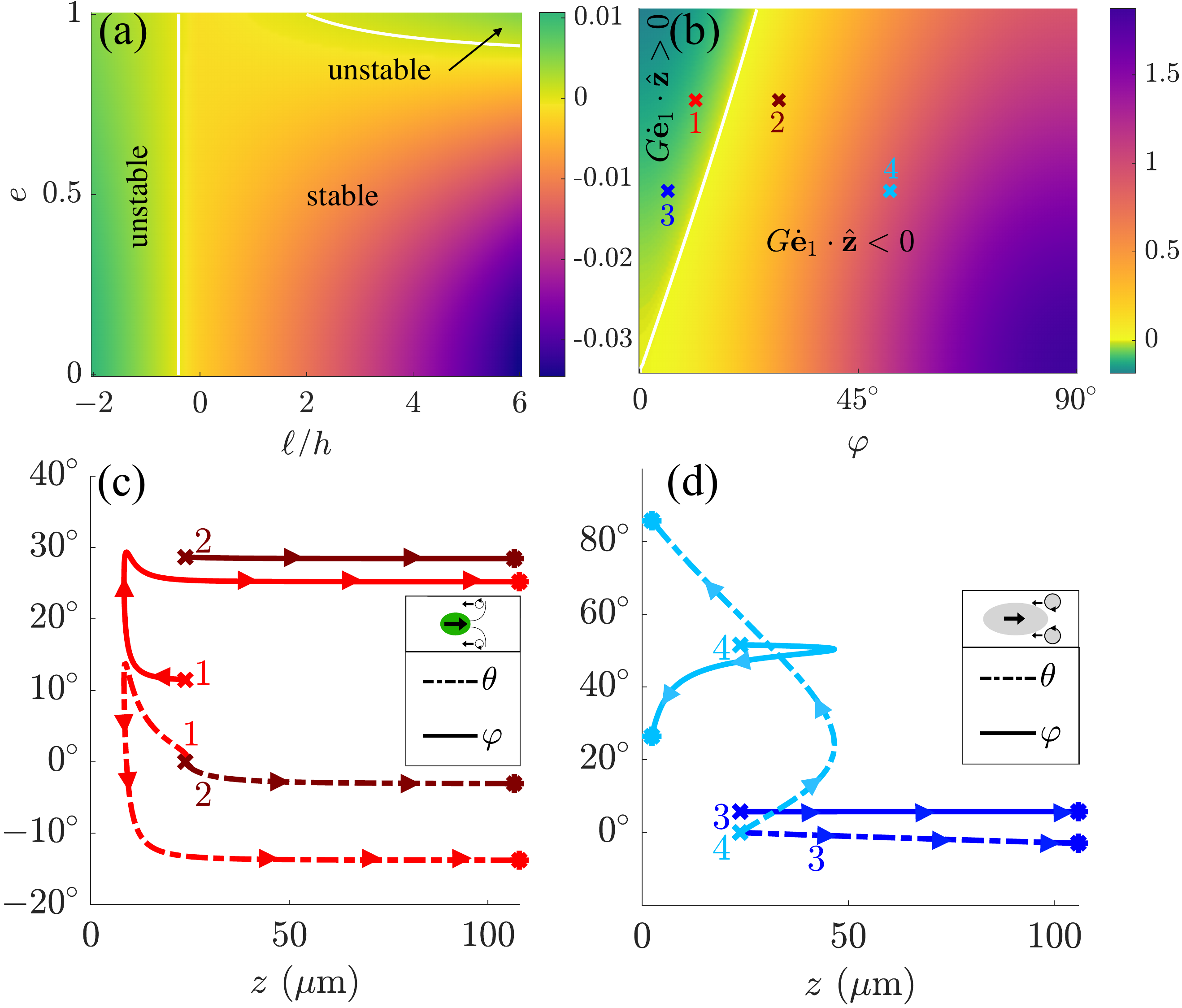}
\caption{Results of stability analysis.  (a) The case of a trajectory initially oriented 
nearly perpendicular to the surface. Trajectories were obtained with the value of $Fd$ extracted 
from PIV, $U=0$, $h=50$ with a nearly vertical swimming direction, $t_{\max}=40$ and a $100\times 100$ grid. 
The aspect ratio $e$ and dimensionless feeder length $\ell/2h$ were varied, and the plot 
shows $\max_i [|e_i(t_{\max})|-|e_i(0)|]$ for $i\in\{1,2\}$. White lines mark the theoretical 
boundaries of the stability regions. (b) Stability diagram of a rotlet dipole with 
$\mathbf e_1(0)=\hat{\mathbf x}$ and roll $\varphi$. For $G>0$ (feeder) the dipole is rotated 
towards the surface ($\dot{\mathbf e}_1\cdot\hat{\mathbf z}<0$) for $\varphi$ sufficiently large, 
while when $G<0$ (expeller) the dipole rotates away from the surface unless $\varphi$ is small. 
(c) Sample trajectories for representative parameter values of \textit{C. reinhardtii} in different 
parts of phase space. (d) Sample trajectories for representative parameter values of rotifer in 
different parts of phase space.}
\label{fig3}
\end{figure*}

Considering now initially parallel trajectories with $\mathbf e_1=\hat{\mathbf x}$, 
$\mathbf e_2=\hat{\mathbf y}\cos\alpha+\hat{\mathbf z}\sin\alpha$, 
$\mathbf e_3=-\hat{\mathbf y}\sin\alpha+\hat{\mathbf z}\cos\alpha$, since the force dipole does not 
rotate the swimmer \cite{Spagnolie}, the leading-order effect comes from the rotlets. The presence of 
the rotlet tends to steer the swimmer towards the wall for feeders and away from the wall for expellers. 
Explicitly, we find that the swimmer is rotated towards the wall if 
\begin{align}
G\left[6\sin^2\varphi+\Gamma \left(1+4\sin^2\varphi\right)-1\right]>0.
\label{Parallel swimming evolution equation} 
\end{align}
This is always the case for feeders provided $\alpha$ is not too close to a multiple of $\pi$. 
If this happens, vorticity prevails over shear alignment and the swimmer turns away from the wall. 
For expellers \eqref{Parallel swimming evolution equation} predicts that the swimmer will tend 
to turn away from the wall provided $\sin \alpha$ is not too small, which matches 
numerical findings \cite{ChlamyScattering}.

The dynamics of trajectories can be quite complex as a consequence of the competition between
the stresslet and rotlet dipole contributions. Figure \ref{fig3} shows four trajectories 
for feeders and expellers initially oriented parallel to the surface. 
Fig.~\ref{fig3}(c) shows two trajectories corresponding to parameters appropriate to \textit{C.~reinhardtii} 
with initial roll $\varphi_0=11.5^{\circ}$ and $\varphi_0=28.6^{\circ}$. For the former, just 
inside the stability region, the pitch initially increases and the cell moves towards the wall, 
until $\varphi$ becomes 
too large. When this occurs, the organism rapidly rotates away from the wall (decrease in $\theta$) and 
then departs, never to return. For $\varphi_0=28.6^{\circ}$, the alga rotates away from the wall and 
departs. Figure \ref{fig3}(d) instead shows two trajectories appropriate to {\it B. plicatilis} with $\varphi_0=5.7^{\circ}$ 
and $\varphi_0=51.6^{\circ}$. For the former, inside the unstable region, the rotifer swims 
away from the wall. When $\varphi_0=51.6^{\circ}$, the organism rotates towards the wall and crashes 
after a temporary increase in $z$ due to the force dipole.

\section{Motility Statistics Near Surfaces}
\label{motility}

\begin{figure*}[t]
\centering
\includegraphics[width=0.8\columnwidth]{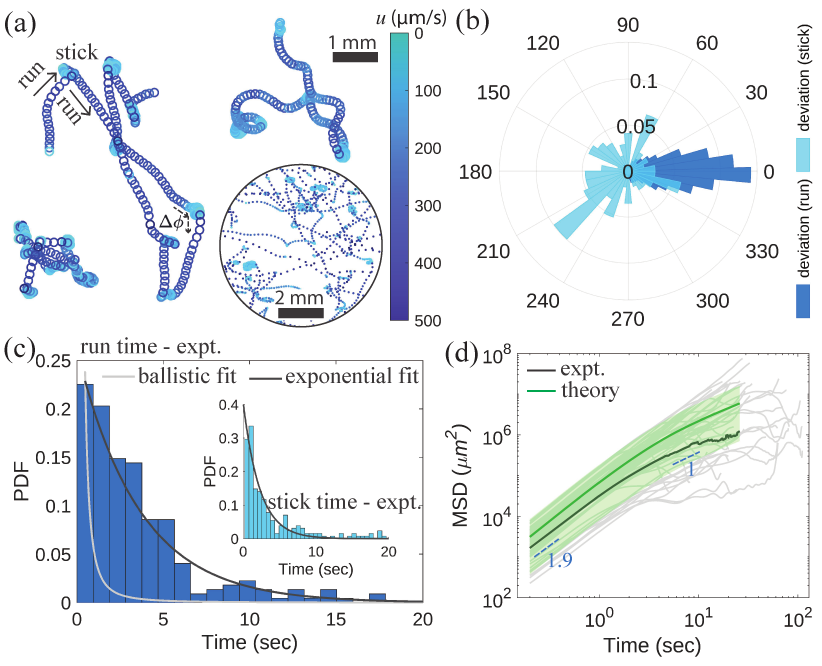}
\caption{Motility statistics of swimming rotifers. (a) Examples of trajectories near 
the top chamber surface, color-coded by the swimming speed. (b) PDF of `run deviations' 
between consecutive stick events and `stick deviations', 
the difference between incoming and outgoing angles. (c) PDF of the run-time and stick-time durations. 
(d) MSD (gray - individual tracks, black - average) of \textit{B. plicatilis} exhibits a transition 
from super-diffusive to diffusive behavior, owing to sticking near the surface. The green curve and 
shading illustrate the predicted MSD and $2$ standard deviations based on theoretical calculations.}
\label{fig4}
\end{figure*}

The transitions back and forth that rotifers exhibit 
between free swimming parallel to a surface and to attaching to it lead to an
intriguing type of random walk loosely analogous to several other multi-mode locomotion 
motifs: the ``run-and-tumble" locomotion of bacteria such
as \textit{E. coli} \cite{BergBrown}, the ``run-reverse-flick" locomotion of 
\textit{Vibrio alginolyticus} \cite{runflickreverse},
and the ``run-and-turn" locomotion of \textit{Chlamydomonas} \cite{Polin}, and
the ``run-stop-shock" locomotion of \textit{Pyramimonas octopus} \cite{octopus}.
It is perhaps most closely related to the behavior exhibited by a particular 
pathogenic strain of \textit{E. coli} that stochastically transitions between
runs parallel to surfaces and adhesive events \cite{Peruani}.
We analyzed $62$ tracks of \textit{B. plicatilis} with typical 
durations ranging of $20\,$s.
Figure \ref{fig4}(a) shows typical examples of those trajectories, which follow 
what we term a ``run-and-stick" sequence.  
As detailed in Sec. \ref{scattering}, a rotifer that is close to a
surface may rapidly reorient to become perpendicular to the surface, becoming 
stably attached, as in Fig.~\ref{fig2}(d). While the rotifer's strong puller stresslet 
flow field leads to strong attachment, they do eventually release owing to biological activity 
(SM Video 3 \cite{SM}). While attached, the aforementioned spinning of rotifers around
the body axis $\mathbf e_1$ leads to a randomized swimming direction upon their release 
(SM Video 1 \cite{SM}). 

We quantify the run-and-stick trajectories by examining first in Fig.~\ref{fig4}(b) 
two measures of angular deviations: (i) the `run-deviations' 
$\Delta \theta_i=\theta_i - \langle\theta_i\rangle$
of trajectory directions $\theta_i$ from the mean direction $\langle\theta_i\rangle$ of the 
$i^{\rm th}$ swimming event, sampled in steps of $0.2\,$s, and (ii) the   
`stick-deviations', the angular change $\Delta\phi$ between incoming and outgoing directions
at each sticking event. A sharp peak around $0^{\circ}$ for run-deviations suggests that the 
cells swim mostly in a straight line until reoriented by stick events. The role of biological 
activity in sticking and unsticking events becomes even more evident once we assess the 
`run time' and `stick time' histograms in Fig.~\ref{fig4}(c). Averaged over all $62$ analyzed 
tracks near the top surface we find $\langle {\rm run\,time} \rangle = 3.6\,$s and 
$\,\langle {\rm stick\,time}\rangle = 5.5\,$s, with both distributions decaying exponentially,
signifying a Poisson processes.  Such a process is consistent with the picture that 
the organism swims roughly parallel to the wall until its pitch stochastically fluctuates
sufficiently to trigger a snapping event.  In this case the waiting time between 
events should be exponentially distributed with probability density function (p.d.f.) 
$p_e(t)= \lambda e^{-\lambda t}$.  An alternate view on the waiting time distribution is
that the organism swims ballistically at all times except close to snapping and that all 
the randomness in the trajectory is embedded in the initial pitch $\theta$.
In this case 
we can evaluate the corresponding `ballistic' impact 
time $t_*$ for an incidence angle $\theta$ as $t_*\sim t_0/\sin\theta$, where $t_0=h/U$ 
is a typical swimming time. Because the transition to sticking occurs rapidly relative to the 
swimming timescale, it can be ignored. If the incidence angle is chosen uniformly at random 
in $[0,\pi/2]$, the `run time' distribution follows the ballistic p.d.f.
$p_b(t)=2t_0/\pi t(t^2-t_0^2)^{1/2}$. 
Going back to Fig.~\ref{fig2}(e), we test this approximation by fitting $p_b$ to the p.d.f. 
of the collision time for $100$ trajectories with uniformly distributed initial pitches 
and an initial height $h=360\ \mu$m (inset). Results show excellent agreement with 
$\chi^2$ error $\sim 0.02$.
 
In order to see which of the exponential or ballistic p.d.f. better accounts for
the data, we fitted both $p_e$ and $p_b$ to the data (without automated binning) 
and computed the $\chi^2$ error in the fit. For $p_e$, this was $\sim 0.15$, while for $p_b$ 
it was $\sim 0.74$. This analysis shows that snapping is explained much more convincingly by 
swimming noise rather than ballistic motion with initially random angles. 
From the exponential fit, we find a mean free-flight time of $\sim 3.6$ s.

Next we examine the mean squared displacement (MSD) of rotifers undergoing run-and-stick locomotion.
Figure \ref{fig4}(d) shows the ensemble of MSD measurements for the analyzed tracks and the 
ensemble average MSD.  There is a clear transition from near-ballistic motion for 
short times to a diffusive regime on longer timescales, with a crossover time of $\sim 2-4\,$s. 
To explain this, we propose a mean-field model of a population of rotifers near a boundary in
which the population is coarse-grained into a local number density of freely swimming 
and trapped rotifers. 

Let $f^+(\mathbf x,\theta,t)$ be the number density of freely swimming rotifers at position 
$\mathbf x$ and angle $\theta$ at time $t$. Similarly, let $f^-(\mathbf x,t)$ be the 
number density of trapped rotifers at position $\mathbf x$ and time $t$. 
Based on the motility statistics in Fig.~\ref{fig4}, we assume that individual rotifers 
transition between freely swimming and trapped at a rate $\nu_-$ and from a trapped state to 
freely swimming at a rate $\nu_+$. Neglecting diffusion between trapping events on account of the 
small angular displacement in Fig.~\ref{fig4}(b), we propose that the distributions obey 
the evolution equations 
\begin{subequations}
\begin{align}
\frac{\partial  f^+}{\partial  t}&+V\mathbf n\cdot\nabla_{\mathbf x} f^+=
-\nu_-f^++\frac{\nu_+}{2\pi}f^-\label{Evolution Equation 1 Main}\\
\frac{\partial  f^-}{\partial  t}&=
-\nu_+f^-+\nu_- \int_0^{2\pi}f^+(\mathbf x,\theta,t)\mathrm d\theta,
\label{Evolution Equation 2 Main}
\end{align}
\end{subequations}
where $\mathbf n=[\cos\theta,\ \sin\theta]$. 
We term \eqref{Evolution Equation 1 Main} and \eqref{Evolution Equation 2 Main} 
a \textit{run-and-stick model}. 
Eq.~\ref{Evolution Equation 1 Main} is the statement that a parcel of freely swimming 
rotifers with angle $\theta$ loses $\nu_-f^+$ swimmers per unit time due to sticking, 
and gains $\nu_+f^- \mathrm d\theta/2\pi$ swimmers per unit time as a result of unsticking events. 
The $1/2\pi$ factor denotes the fact that the swimming angle upon release is uniformly random 
due to loss of orientation in the trapped state. Likewise, Eq.~\ref{Evolution Equation 2 Main} models the fact that the trapped population loses members at a rate $\nu_+$ and gains members 
(with arbitrary swimming angle) at a rate $\nu_-$. From $f^+$ and $f^-$ we 
may define the total rotifer number density $\rho(\mathbf x,t)$ as
\begin{align}
\rho(\mathbf x,t)= f^-(\mathbf x,t)+\int_0^{2\pi}f^+(\mathbf x,\theta,t)\mathrm d\theta, 
\label{rho definition} 
\end{align}
which, from \eqref{Evolution Equation 1 Main} and \eqref{Evolution Equation 2 Main}, 
satisfies the conservation law
\begin{align}
\frac{\partial \rho}{\partial t}+\nabla_{\mathbf x}\cdot\int_{0}^{2\pi}V\mathbf n f^+\mathrm d\theta=0.
\end{align}

Appendix \ref{app_d} provides the details of the calculation of the MSD of the
run-and-stick model, with the result
\begin{align}
\frac{\langle r^2(t)\rangle}{2V^2}=
\frac{\nu_-^2 -\nu_- \nu_+ -\nu_+^2 }{\nu_-^2 {{\left(\nu_- +\nu_+ \right)}}^2 }
+\frac{\nu_+ t}{\nu_-{\left(\nu_- +\nu_+ \right)}}
+\frac{\nu_- e^{-t{\left(\nu_- +\nu_+ \right)}} }{\nu_+ {{\left(\nu_- +\nu_+ \right)}}^2 }
-\frac{e^{-\nu_-t} {\left(\nu_- -\nu_+ \right)}}{\nu_-^2 \nu_+ }.
\label{MSD Exact}
\end{align}
At short times $t\ll \nu^{-1}_-$, there is ballistic behavior with 
$\langle r^2(t)\rangle=V^2t^2+\mathcal O(V^2\nu_-t^3)$, 
crossing over to linear behavior at long times, from which we calculate an effective
diffusion coefficient for the 
population based on the asymptotic result $\text{MSD}\sim 4Dt$ for $t\to\infty$ in 
two dimensions (Appendix \ref{app_d}), yielding
\begin{align}
D=\frac{V^2\nu_+}{2\nu_-(\nu_-+\nu_+)}.
\label{Effective Diffusion Coefficient}
\end{align}
Taking $V=280\ \mu \text{m}\ \text{s}^{-1}$, $\nu_+=(5.5\ \text{s})^{-1}$, $\nu_-=(3.6\ \text{s})^{-1}$, 
we obtain $D\sim 0.057\ \text{mm}^2\ \text{s}^{-1}$, in good agreement with experiments.

The result \eqref{Effective Diffusion Coefficient} can be compared with the classic 
run-and-tumble (RT) process,
which formally corresponds to the limit $\nu_+\to\infty$.  This yields the
quasi-steady result $\partial f^-/\partial t=0$, and hence
\begin{align}
\frac{\partial  f^+}{\partial  t}+V\mathbf n\cdot\nabla f^+=\frac{\nu_-}{2\pi}\int_0^{2\pi}f^+\mathrm d\theta-\nu_-f^+,
\end{align}
which corresponds to an RT process with frequency $\nu_-$ \cite{Diffusion Paper}.
In this limit, $D\nearrow D_{\text{RT}}=V^2/2\nu_-$; 
for finite $\nu_+$, $D<D_{\text{RT}}$ so the population 
spreads more slowly than for RT due to the extra latency from sticking events.
Equivalently, there an effective free-flight time smaller than the inverse sticking rate,
 \begin{align}
 \tau_{\text{eff}}=\frac{\nu_+}{\nu_-(\nu_-+\nu_+)}<\frac{1}{\nu_-}. 
 \end{align}

\section{Discussion}
\label{discussion}

In this paper, we demonstrated that puller microswimmers may be classified as  ``feeders'' or  
``expellers'' depending on the sense of circulation of the cilia-driven flows. These two classes 
may be identified by their near-field flows (Fig.~\ref{fig1}), with expellers exhibiting a 
stagnation point ahead of the cell body and feeders conversely presenting incoming flow at 
their apex. Boundary interactions also differ in relation to the ``feeder number'' Fe quantifying 
the relative strengths of the Stokeslet and rotlets associated with propulsion. While strong 
expellers ($|\text{Fe}|\gg 1$) glide along the wall and then depart, feeders tend to collide 
and even attach to the boundary. By comparing with experimental data, we demonstrated that 
such behaviour is closely captured by an approximation consisting of a force dipole and a rotlet 
dipole located at the same point within the organism body (Fig.~\ref{fig2}). 
A linear stability analysis confirms that the rotlet dipole generally turns feeders towards the boundary 
for both parallel and orthogonal incoming trajectories, leading to collision, while expellers 
are rotated away from the boundary, leading to scattering (Fig.~\ref{fig3}). Motility 
statistics (Fig.~\ref{fig4}) reveal that both the free-flight and the trapped times of 
rotifers are exponentially distributed, signifying that sticking and unsticking are well 
described by Poisson processes. While rotifers perform nearly ballistic motion on timescales 
much shorter than the average free-flight time, their long-term motion is significantly 
modified by sticking events. In particular, the changes of incoming and outgoing directions 
are random, leading to a crossover from ballistic to diffusive scaling. 
The motility near the surface is well-described by a mean field run-and-stick model predicting 
an effective diffusive behaviour in good quantitative agreement with experimental data. 

While the results presented here show that many aspects of the swimming dynamics of rotifers 
can be understood using familiar methods in fluid mechanics, we note that 
rotifers possess muscles with which they can
deform their body and sensory organs for touch and light.  They are therefore capable 
of significantly more complex behaviors than the bacteria and protists that serve as
paradigms of microswimmers.  Thus, their dynamics near 
surfaces may reflect at least in part a tactic response as much as a purely passive 
hydrodynamic phenomenon.  Finally, the possibility of interesting collective effects,
whether in bulk or at surfaces, from organisms such as rotifers remains to be explored.

\begin{acknowledgments}
We thank Rebecca Poon and Francesco Boselli for assistance with PIV, Kyriacos Leptos for 
numerous discussions and Brian Ford for historical background on rotifers.  
This work was supported in part
by Grant No. 7523 from the Gordon and Betty Moore Foundation, and the John Templeton 
Foundation. 
\end{acknowledgments}


\appendix

\section{Fitting the flow}
\label{app_a}

The rotifer's body is seen to lie within the focal plane throughout the analyzed videos, 
implying that it is parallel to the no-slip wall at $z=0$. 
We thus orient the PIV field of view so that the body frame of reference is 
$\{\mathbf e_1,\mathbf e_2,\mathbf e_3\}=\{\hat{\mathbf x},\hat{\mathbf y},\hat{\mathbf z}\}$ and 
fit the flow in the $xy$ plane. Such procedure is performed after removing the solid-body motion 
velocity corresponding to the rotifer location. We model the rotifer via a far-field approach 
whereby the body and locomotion apparatus are replaced by point singularities; 
the thrust is taken to be being parallel to the body axis, and the balancing effects of the cilia 
bundle and the body drag are represented by Stokeslets of strength $F\hat{\mathbf x}$, $-F\hat{\mathbf x}/2$, 
$-F\hat{\mathbf x}/2$ located at $\mathbf x_0$, $\mathbf x_{\pm}=\mathbf x_0+d\hat{\mathbf x}
\pm w\hat{\mathbf y}$. We additionally place two rotlets of strengths $\mp G\mathbf e_3$ at 
$\mathbf x_{\pm}$, as in Fig. \ref{fig2}. Because the field of view is 
confined to the $xy$ plane, we can only detect the $\hat{\mathbf z}$ component of the rotlets from 
the cilia bundles. The flow Ansatz is then given by Eq.~\ref{FittingAnsatz} in the main text, where 
$\mathbf B$ is the Green's function for a point force near a no-slip boundary,
\begin{equation}
\mathbf B_{ij}(\mathbf x;\mathbf x_0)=\left(\frac{\delta_{ij}}{r}+\frac{r_ir_j}{r^3}\right)-
\left(\frac{\delta_{ij}}{R}+\frac{R_iR_j}{R^3}\right)+2h(\delta_{j\beta}\delta_{\beta k}-
\delta_{j3}\delta_{3k})\frac{\partial }{\partial  R_k}\left[\frac{h R_i}{R^3}-\left(\frac{\delta_{i3}}
{R}+\frac{R_iR_3}{R^3}\right)\right],
\label{Blake Stokeslet in components}
\end{equation}
with $\beta=1,2$, $h=\mathbf x_0\cdot\hat{\mathbf z}$, $\mathbf r=\mathbf x-\mathbf x_0$, 
$\mathbf R=\mathbf x-\mathbf x_0^*$ and $\mathbf x_0^*=\mathbf x_0-2h\hat{\mathbf z}$. 
Similarly, $\mathbf R$ is the Green's function for a point torque near a no-slip boundary,
\begin{equation}
\mathbf R_{ij}(\mathbf x;\mathbf x_0)=\frac{\varepsilon_{ijk}r_k}{r^3}-\frac{\varepsilon_{ijk} R_k}{R^3}
+2h\varepsilon_{kj3}\left(\frac{\delta_{ik}}{R^3}-\frac{3R_iR_k}{R^5}\right)
+6\varepsilon_{kj3}\frac{R_iR_kR_3}{R^5}.
\label{Blake rotlet in components}
\end{equation}
For simplicity, we pin $\mathbf x_0$ on 
the body axis, so we only fit the distance from the wall and the axial position of the body Stokeslets, 
giving $6$ fitting parameters in total.

In fitting the flow field, we care particularly about capturing the four near-field lobes, a signature 
of proximity to a no-slip surface. We therefore propose as the metric to minimize the function
\begin{align}
\lVert \mathbf u_p-\mathbf u_a\rVert_{\star}= \frac{1}{2}\sum_{\mathbf x_i}(1-\cos\theta_i)w(\mathbf x),
\end{align}
where $\mathbf u_p$ and $\mathbf u_a$ are the PIV and analytical flows, $\theta_i$ is the angle with 
respect to the PIV flow direction and the $\mathbf x_i$ are the PIV lattice nodes after removal of 
the region corresponding to the rotifer's body. The weight function $w(\mathbf x)$ is the sum of 
four Gaussians centred at the vortices, 
\begin{equation}
w(\mathbf x)=C\sum_{n=1}^4 \exp\left(-\frac{\lVert\mathbf x-\mathbf x_i\rVert^2}{2\sigma^2}\right),    
\end{equation}
where $\sigma$ is the standard deviation (set by the vortex size), assumed for simplicity to be 
equal for all vortices, and $C$ is a (numerically determined) normalising constant such that
\begin{align}
\sum_{\mathbf x_i}w(\mathbf x_i)=1.
\end{align}
We chose $\sigma=1$ in our fitting and constrained the singularity locations lie close to the 
solid-body region of PIV, specifically within a rectangle $1.5$ times larger than the rotifer's body.
This fit only determines the flow up to rescaling $f\to \kappa f$, $G\to \kappa G$. We find the optimal value 
of $\kappa$ via a least-squares method.

\section{Equations of Motion}
\label{app_b}
In the absence of a wall, the swimmer swims without rotation in a straight line with velocity $U\mathbf e_1$.
The no-slip condition on the wall induces a perturbation flow $\mathbf u^*(\mathbf x)$ that 
may formally be obtained by placing suitable ``image singularities'' at the mirror-image of the body's 
location $\mathbf x_0^*=\mathbf x_0-2(\mathbf x_0\cdot\hat{\mathbf z})\hat{\mathbf z}$ in order to 
exactly cancel the organism's flow at the wall. Such a flow advects and rotates the organism according to 
the Faxén laws for a prolate ellipsoid,
\begin{align}
&\dot{\mathbf x}_0=U\mathbf e_1+\mathbf u^*(\mathbf x_0)\label{x_0 dot equation}\\
&\frac{\mathrm d}{\mathrm dt}\{\mathbf e_1,\mathbf e_2,\mathbf e_3\}=\left[\frac{1}{2}\nabla 
\times\mathbf u^*(\mathbf x_0)+\Gamma\mathbf e_1\times \mathbf E^*(\mathbf x_0)\cdot\mathbf e_1\right]
\times \{\mathbf e_1,\mathbf e_2,\mathbf e_3\},
\label{Rotation Equation}
\end{align}
where $\Gamma=(1-e^2)/(1+e^2)$ is the Bretherton parameter encoding shear-alignment, 
$0\leq e\leq 1$ is the ellipsoid's aspect ratio, and 
$\mathbf E^*=(\nabla \mathbf u^*+\nabla^{\text T}\mathbf u^*)/2$ is the rat-of-strain tensor of the image flow. 
For simplicity, we henceforth model the rotifer as a superposition of a force dipole of strength $fd$ 
along $\mathbf e_1$ and a rotlet dipole of strength $Gw$ along $\mathbf e_2$.

The image flow $\mathbf u^*_{\text{rd}}$ induced by the rotlet dipole (located at $\mathbf x_0$) 
at an arbitrary point $\mathbf x$ is
\begin{align}
\mathbf u^*_{\text{rd}}(\mathbf x)&=-\frac{wG}{4\pi\mu}\mathbf e_2\cdot\nabla_{\mathbf y}
\mathbf R^*(\mathbf x;\mathbf y)\lvert_{(\mathbf x;\mathbf x_0)}\cdot \mathbf e_3.
\label{Rotlet dipole velocity}
\end{align}
The image flow $\mathbf R^*(\mathbf x;\mathbf y)\cdot\mathbf e_3$ generated by a rotlet of strength 
$\mathbf e_3$ located at $\mathbf y$ is
\begin{align}
\mathbf R^*_{ij}(\mathbf x;\mathbf x_0)&=-\frac{\varepsilon_{ijk} R_k}{R^3}+2(\hat{\mathbf z}
\cdot \mathbf y)\varepsilon_{kj3}\left(\frac{\delta_{ik}}{R^3}-\frac{3R_iR_k}{R^5}
\right)+6\varepsilon_{kj3}\frac{R_iR_kR_3}{R^5},
\label{R star definition}
\end{align}
where $\mathbf R=\mathbf x-\mathbf y^*$, $\mathbf y^*=\mathbf y-2(\hat{\mathbf z}\cdot\mathbf y)\hat{\mathbf z}$. 
The vorticity and rate-of-strain tensor of this flow at the position $\mathbf x_0$ are
\begin{align}
\nabla \times\mathbf u_{\text{rd}}^*(\mathbf x_0)&=-\frac{wG}{4\pi\mu}\nabla_{\mathbf x}\times
\left[\mathbf e_2\cdot\nabla_{\mathbf y}\mathbf R^*(\mathbf x;\mathbf y)
\cdot \mathbf e_3\right]\lvert_{(\mathbf x_0;\mathbf x_0)}\label{Rotlet dipole vorticity},\\   
\mathbf E^*_{\text{rd}}(\mathbf x_0)&= -\frac{wG}{8\pi\mu}\nabla_{\mathbf x}
\left[\mathbf e_2\cdot\nabla_{\mathbf y}\mathbf R^*(\mathbf x;\mathbf y)\cdot \mathbf e_3\right]-
\frac{wG}{8\pi\mu}\nabla_{\mathbf x}^{\text T}\left[\mathbf e_2\cdot\nabla_{\mathbf y}
\mathbf R^*(\mathbf x;\mathbf y)\cdot \mathbf e_3\right]\lvert_{(\mathbf x_0;\mathbf x_0)}.
\label{Rotlet dipole rate-of-strain}
\end{align}
On the other hand, the image flow $\mathbf u^*_{\text{fd}}$ produced by the force dipole 
(located at $\mathbf x_0$) at an arbitrary point $\mathbf x$ is
\begin{align}
\mathbf u^*_{\text{fd}}(\mathbf x)&=-\frac{fd}{8\pi\mu}\mathbf e_1\cdot\nabla_{\mathbf y}
\mathbf B^*(\mathbf x;\mathbf y)\lvert_{(\mathbf x;\mathbf x_0)}\cdot\mathbf e_1,
\label{Force dipole velocity}
\end{align}
where the image flow $\mathbf B^*(\mathbf x;\mathbf y)\cdot \mathbf e_1$ generated by a Stokeslet 
of strength $\mathbf e_1$ located at $\mathbf y$ is 
\begin{equation}
\mathbf B^*_{ij}(\mathbf x;\mathbf y)=-\left(\frac{\delta_{ij}}{R}+\frac{R_iR_j}{R^3}\right)
+2(\hat{\mathbf z}\cdot \mathbf y)(\delta_{j\beta}\delta_{\beta k}
-\delta_{j3}\delta_{3k})\frac{\partial }{\partial  R_k}\left[\frac{h R_i}{R^3}
-\left(\frac{\delta_{i3}}{R}+\frac{R_iR_3}{R^3}\right)\right],
\label{G star definition}
\end{equation}
where $\mathbf R$ is defined as in \eqref{R star definition}. The corresponding 
vorticity and rate-of-strain tensor are
\begin{align}
\nabla \times\mathbf u^*_{\text{fd}}(\mathbf x_0)&=-\frac{fd}{8\pi\mu}\nabla_{\mathbf x}
\times\left[\mathbf e_1\cdot\nabla_{\mathbf y}\mathbf B(\mathbf x;\mathbf y)
\cdot\mathbf e_1\right]\lvert_{(\mathbf x_0;\mathbf x_0)},
\label{Force dipole vorticity}\\
\mathbf E^*_{\text{fd}}(\mathbf x_0)&=-\frac{fd}{16\pi\mu}\nabla_{\mathbf x}\left[\mathbf e_1
\cdot\nabla_{\mathbf y}\mathbf B(\mathbf x;\mathbf y)\cdot\mathbf e_1\right]-\frac{fd}{16\pi\mu}
\nabla_{\mathbf x}^{\text T}\left[\mathbf e_1\cdot\nabla_{\mathbf y}\mathbf B(\mathbf x;\mathbf y)
\cdot\mathbf e_1\right]\lvert_{(\mathbf x_0;\mathbf x_0)}.
\label{Force dipole rate-of-strain}
\end{align}
The combined effect of both singularities is obtained by adding up their respective contributions 
to the flow by linearity. In the numeric studies, we integrate \eqref{x_0 dot equation}) and 
\eqref{Rotation Equation}) for the trajectory with the Matlab \texttt{ode45} routine.

\section{Linear Stability Analysis}
\label{app_c}
\begin{figure*}[t]
\centering
\includegraphics[width=0.6\columnwidth]{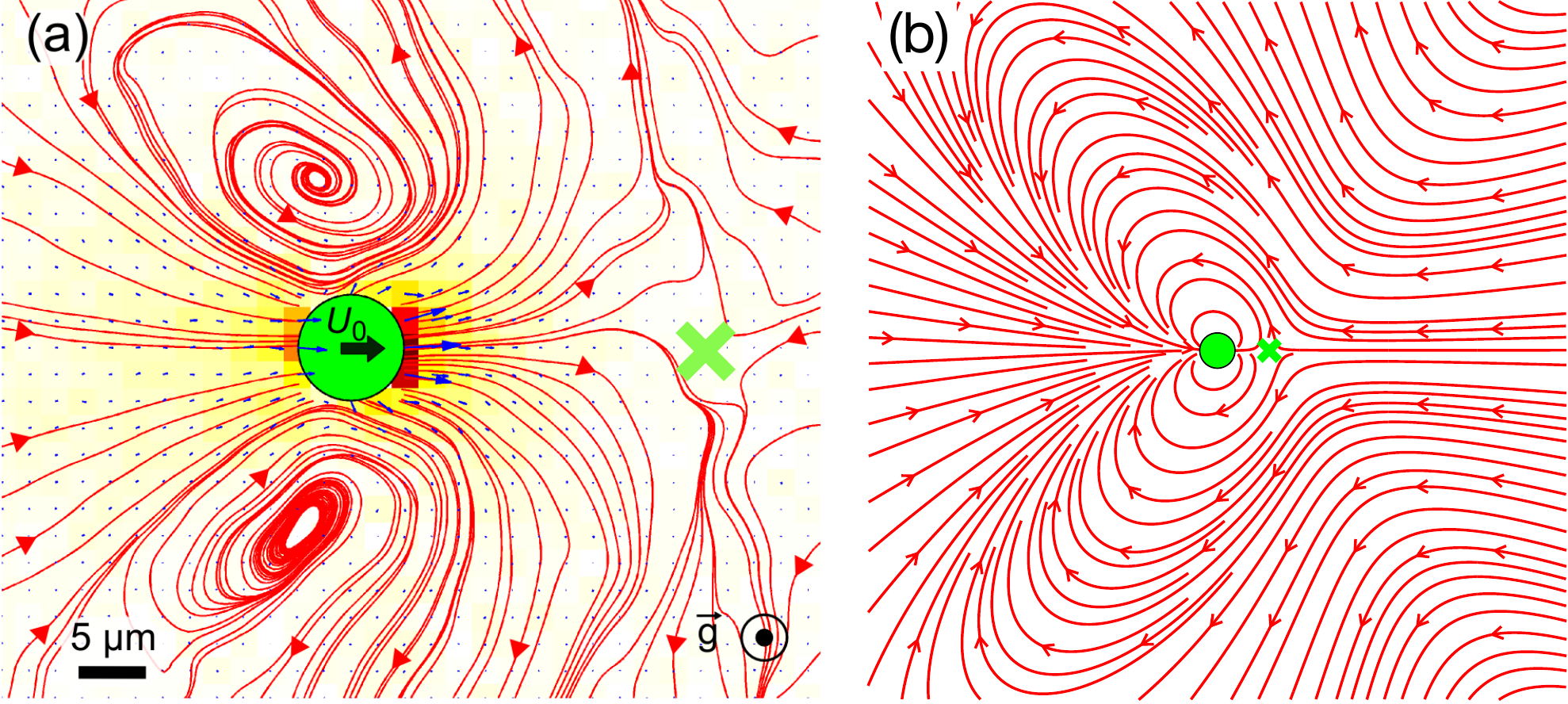}
\caption{Test of the far-field model. (a) Experimental flow field around \textit{C. reinhardtii}, 
from \cite{Direct}. (b) Approximation consisting of two singularities located at the cell-center, 
a force dipole and a rotlet dipole. This differs from the approach in \cite{Direct} in which 
three separated singularities were used.}
\label{fig5}
\end{figure*}

We aim to understand analytically the effect of the far-field singularities on 
the trajectory of a swimmer of typical size $L$ when swimming (nearly) perpendicular or 
parallel to a no-slip surface. Such singularities consist of a force dipole of strength $fd$, 
a rotlet dipole of strength $Gw$, and terms smaller by a factor $\mathcal O(L/\lVert\mathbf x\rVert)$, 
which we neglect. Despite the rotlet dipole being asymptotically smaller, it becomes 
comparable to the force dipole at a distance $\lVert\mathbf x\rVert=\ell\sim Gw/fd$ from the body. 
This implies that for ``strong feeders/expellers'' with $|\ell|\gg L$, the rotlets govern the 
impact dynamics. Indeed, the singularity description is still accurate for such an organism 
as distances from the wall in the range $L\ll z\ll |\ell|$. 
Figure \ref{fig5} shows that such singularities convincingly capture the flow around 
\textit{C.~reinhardtii}, including the stagnation point. This motivates using the above 
far-field description of pullers in the calculations.

\subsection{Dynamics for a Nearly-Perpendicular Trajectory}
\label{app_c1}
A swimmer initially moving perpendicularly to the wall, i.e.~with 
$\{\mathbf e_1,\mathbf e_2,\mathbf e_3\}=\{-\hat{\mathbf z},\hat{\mathbf y},\hat{\mathbf x}\}$,
will continue to move perpendicularly by symmetry. In this section we analyze the evolution 
of a small perturbation
\begin{equation}
\{\mathbf e_1,\mathbf e_2,\mathbf e_3\}=\mathbf \Pi(t)\cdot\{-\hat{\mathbf z},\hat{\mathbf y},\hat{\mathbf x}\},
\end{equation}
for $t\geq 0$, where $\mathbf \Pi_{ij}(t)=\mathbf I_{ij}+\varepsilon\epsilon_{ikj}\mathbf V_k(t)
+\mathcal O(\varepsilon^2)$, with $|\varepsilon|\ll 1$, is an infinitesimal rotation matrix. 
All equations are up to $\mathcal O(\varepsilon^2)$, since $\{\mathbf e_1,\mathbf e_2,\mathbf e_3\}$ must 
have unit length. 
From (\ref{Rotlet dipole velocity}) and (\ref{Force dipole velocity}) we obtain the leading-order velocity
\begin{equation} 
\dot{\mathbf x}_0=-\left[U+\frac{3fd}{32\pi \mu z_0^2}+\frac{Gw}{8\pi\mu z_0^3}\right]\hat{\mathbf z}
+\mathcal O(\varepsilon).
\end{equation}
Thus, feeders ($G>0$) receive a boost from the suction flow setup by the rotlets, 
while expellers ($G<0$) are slowed down. If $G$ is large and negative, eventually 
$\dot{\mathbf x}_0=\mathbf 0$ at leading order, so the organism hovers above the wall.
Turning our attention to the rotational dynamics, from (\ref{Rotlet dipole vorticity}), 
\eqref{Rotlet dipole rate-of-strain}, \eqref{Force dipole vorticity} and 
\eqref{Force dipole rate-of-strain} the rate of turning $\dot{\mathbf e}_3$ is 
\begin{equation}
\dot{\mathbf e}_1=-\frac{3{\left(7\Gamma Gw -Gw +2fdz_0 
+2\Gamma fdz_0 \right)}}{128\pi\mu {z_0 }^4 }(\mathbf e_1\cdot\hat{\mathbf x})\hat{\mathbf x}-
\frac{3{\left(\Gamma +1\right)}{\left(5Gw +2fdz_0 \right)}}{128\pi\mu{z_0 }^4 }(\mathbf e_1
\cdot\hat{\mathbf y})\hat{\mathbf y}.
\label{Perpendicular swimming evolution equation}
\end{equation}
The rotifer is thus rotated towards the wall when $\dot{\mathrm e}_x/{\mathrm e}_x<0$, 
$\dot {\mathrm e}_y/{\mathrm e}_y<0$, which is the case when
\begin{equation}
(7\Gamma-1)Gw+2fdz_0(1+\Gamma)>0, \ \ \ \ {\rm and} \ \ \ \ 5Gw +2fdz_0 >0.
\label{Stability conditions}
\end{equation}
Equation (\ref{Stability conditions}) shows that, unless the body is very nearly spherical, 
strong expellers are rotated away from the wall while strong feeders are rotated towards the wall. 
Rotation towards the wall is facilitated by shear alignment and impeded by the vorticity. 
Indeed, when $\Gamma\ll 1$, $\dot{\mathrm e}_x/\mathrm{e}_x>0$ for $G$ large and positive, 
while if $\Gamma\sim 1$ both $\mathrm e_x$ and $\mathrm e_y$ shrink over time. 
For rotifers, $\Gamma\sim 0.9>1/7$, so the rotlets have a stabilizing effect.
\begin{figure*}[t]
\centering
\includegraphics[width=0.6\columnwidth]{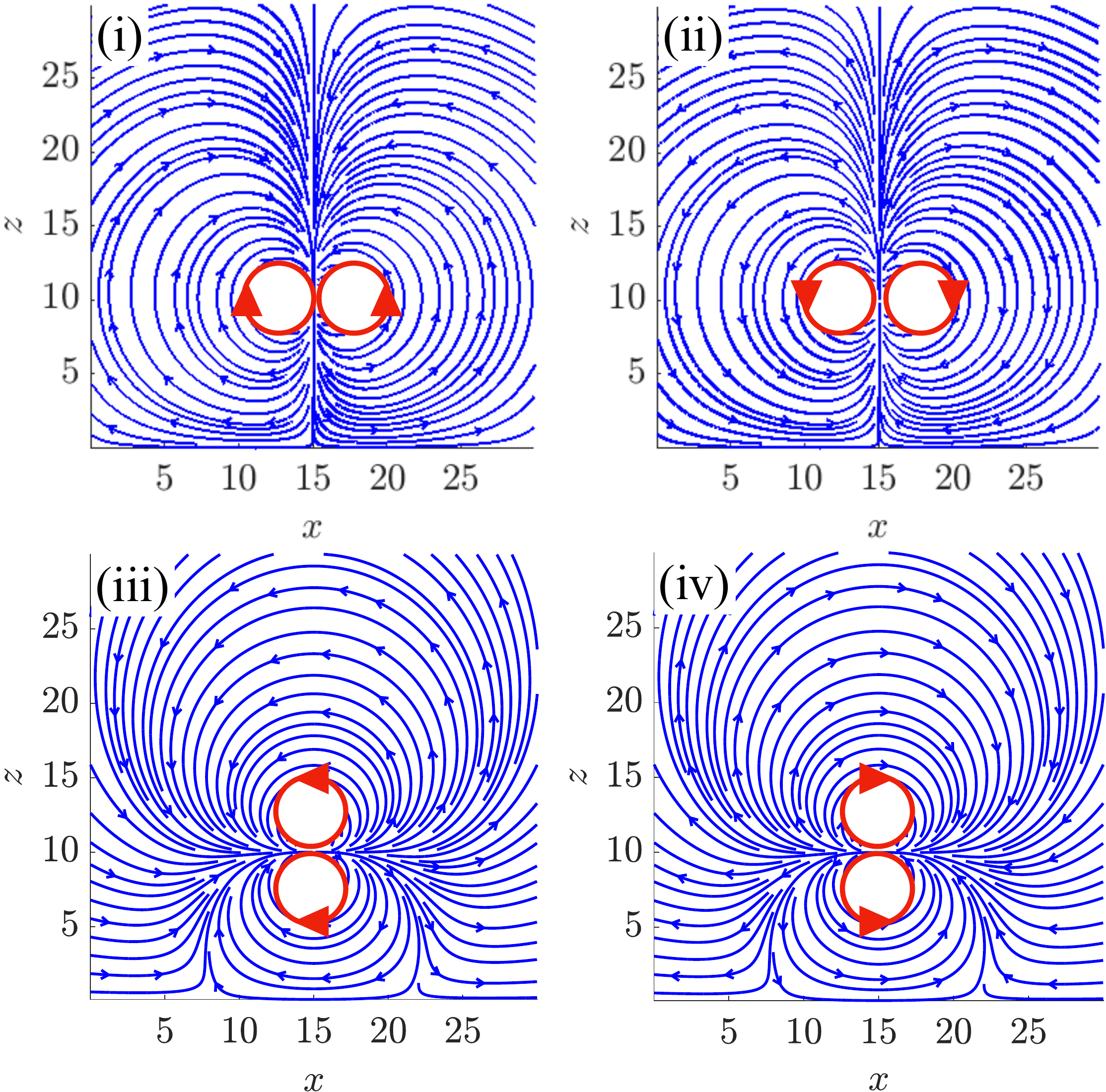}
\caption{Flow produced by a rotlet dipole above a no-slip wall at $z=0$: 
(i) Expeller flow when the swimmer is facing towards the wall; 
(ii) Feeder flow when the swimmer is facing towards the wall; 
(iii) Expeller flow when the swimmer is parallel to the wall; 
(iv) Feeder flow when the swimmer is parallel to the wall}
\label{fig6}
\end{figure*}
\subsection{Dynamics for a Parallel Trajectory}
\label{app_c2}
Similarly to section \ref{app_c1}, we compute the translational and orientational dynamics for 
swimming parallel to the surface, i.e.~when
\begin{equation}
\mathbf e_1=\hat{\mathbf x}, \ \ \  \mathbf e_2=\hat{\mathbf y}\cos\alpha+\hat{\mathbf z}\sin\alpha, 
\ \ \ \mathbf e_3=-\hat{\mathbf y}\sin\alpha+\hat{\mathbf z}\cos\alpha,
\label{Parallel Trajectory Initial Configuration}
\end{equation}
where $\alpha\in [0,\pi/2]$ is the ``roll'' angle around the body axis. 
Unlike in \ref{app_c1}, the material frame \eqref{Parallel Trajectory Initial Configuration} 
rotates even in the absence of an initial perturbation. From \eqref{Rotlet dipole velocity} and 
\eqref{Force dipole velocity} we obtain the leading-order instantaneous translational velocity, 
while \eqref{Rotlet dipole vorticity}, \eqref{Rotlet dipole rate-of-strain}, \eqref{Force dipole vorticity}
and \eqref{Force dipole rate-of-strain} provide the instantaneous rate of turning towards the wall
\begin{align}
&\dot{\mathbf x}_0=\left(U+\frac{Gw}{32\pi\mu z_0^3}\right)\hat{\mathbf x}
+\frac{3fd}{64\pi\mu z_0^2}\hat{\mathbf z}\\
&\dot{\mathbf e}_1=-\frac{3 (\Gamma+3)Gw\sin 2\alpha}{256\pi \mu z_0^4}\hat{\mathbf y}
+\left[\frac{3wG{\left(1-6\sin^2\alpha\right)}}{128\pi\mu z_0^4}
-\frac{3w\Gamma G\,{\left(1+4\sin^2\alpha\right)}}{128\pi\mu z_0^4 }\right]\hat{\mathbf z}\\
&\dot{\mathbf e}_2=
\frac{3\left(\Gamma +1\right)Gw\sin \alpha\left(3\sin^2\alpha+2\right)}{128\pi\mu z_0^4}\hat{\mathbf x}\\
&\dot{\mathbf e}_3=
\frac{3Gw \cos\alpha\left(4\Gamma -3\cos^2\alpha -3\Gamma \cos^2\alpha +2\right)}{128\pi\mu z_0^4}\hat{\mathbf x}
\end{align}
As expected, the ``puller'' dipole is repelled by the surface, but it is not rotated by the image flow. 
The rotation is instead driven entirely by the rotlet dipole flow. A feeder ($G>0$) 
experiences a speed boost along the swimming direction, and is rotated towards the wall 
($\dot{\mathbf e}_1\cdot\hat{\mathbf z}<0$) provided that 
\begin{equation}
\sin\alpha >\left(\frac{1-\Gamma}{6+4\Gamma}\right)^{1/2},
\end{equation}
where the right-hand-side ranges from $0$ (when $e=0$) to $6^{-1/2}\sim 0.41$ (when $e=1$).
As in the orthogonal case, rotation towards the wall is promoted by shear 
alignment and, when $\sin \alpha \ll 1$, is impeded by the vorticity. 
Therefore, for most roll angles feeders are attracted to the wall and expellers are repelled by the wall.

The hydrodynamic mechanism for reorientation is explained by examining the flow streamlines, 
plotted in Fig.~\ref{fig6} for strong expellers and feeders. Feeders facing towards the wall 
are sucked in, while expellers are slowed down by the outgoing rotlet flow. As for swimmers 
oriented parallel to the wall, the puller dipole tends to rotate the swimmer towards the wall 
in both case, but the rotlets aid rotation for feeders and impede it for expellers.

\section{Continuum Run-and-Stick Process}
\label{app_d}

In order to solve \eqref{Evolution Equation 1 Main} and \eqref{Evolution Equation 2 Main} for a 
function $g(\mathbf x,t)$ we define the Laplace-Fourier transform (LFT) $\tilde g(\mathbf k,s)$ by
\begin{align}
\tilde g(\mathbf k,s)\coloneqq\int_0^{\infty}\mathrm dt\ e^{-st}
\int_{\mathbb R^2}\mathrm d^2\mathbf x\ e^{-\mathrm i\mathbf k\cdot\mathbf x} g(\mathbf x,t).  
\end{align}
If we assume the initial conditions $f^+(\mathbf x,\theta,0)=\delta^{(2)}(\mathbf x)/2\pi$, 
$f^-(\mathbf x,t)=0$, corresponding to all rotiers released from the origin with uniformly 
random orientations, taking the LFT of \eqref{Evolution Equation 1 Main} and 
\ref{Evolution Equation 2 Main} gives
\begin{align}
&2\pi\left(s +\mathrm iV\mathbf k\cdot\mathbf n+\nu_-\right)\tilde f^+=1+\nu_+\tilde f^-,
\label{Evolution Equation 1 Transform}\\
&(s+\nu_+)\tilde f^-=\nu_-\int_0^{2\pi}\tilde f^+(\mathbf k,\theta,s)\mathrm d\theta.
\label{Evolution Equation 2 Transform}
\end{align}
Using \eqref{Evolution Equation 2 Transform} to eliminate $\tilde f^-$ in 
\eqref{Evolution Equation 1 Transform} and integrating over $\theta$ we obtain
\begin{align}
\int_0^{2\pi}\tilde f^+\mathrm d\theta=
\frac{1}{2\pi}\left(1+\frac{\nu_+\nu_-}{s+\nu_+}\int_0^{2\pi}\tilde 
f^+\mathrm d\theta\right)\int_0^{2\pi}\frac{\mathrm d\theta}{s+\nu_-+\mathrm iV\mathbf k\cdot\mathbf n}.
\end{align}
The integral may be evaluated by letting $z=e^{\mathrm i\theta}$ and using the residue theorem,
yielding
\begin{align}
\int_0^{2\pi}\tilde f^+\mathrm d\theta=\left\{[(s+\nu_-)^2+V^2k^2]^{1/2}
-\frac{\nu_+\nu_-}{s+\nu_+}\right\}^{-1},
\end{align}
where $k= \lVert\mathbf k\rVert$. We may now evaluate $\tilde f^{-}$ from 
\eqref{Evolution Equation 2 Transform} and express the LFT of the total number 
density as
\begin{align}
\tilde \rho= \frac{s+\nu_++\nu_-}{s+\nu_+}\left\{[(s+\nu_-)^2+V^2k^2]^{1/2}
-\frac{\nu_+\nu_-}{s+\nu_+}\right\}^{-1}.
\label{tilde rho expression}
\end{align}
Denoting the Laplace transform by $\mathcal L$, we may exploit the rotational symmetry of $\rho$ 
to write the MSD directly in terms of $\tilde \rho$,
\begin{align}
\langle r^2(t)\rangle=2\pi\int_{0}^{\infty}\rho(r,t)r^3\mathrm dr
=\mathcal L^{-1}\left\{-2\left.\frac{\partial^2}{\partial k^2} \tilde\rho(k,s)\right|_{k=0}\right\},
\label{MSD from rho tilde}
\end{align}
Evaluating the inverse Laplace transform by means of \eqref{tilde rho expression}, we obtain 
the analytical expression \eqref{MSD Exact} for the MSD given in the main text.

\end{document}